# Updating RDFS ABoxes and TBoxes in SPARQL

Albin Ahmeti[3], Diego Calvanese[2], and Axel Polleres[1]

[1] Wirtschaftsuniversität Wien (WU), Welthandelsplatz 1, 1020 Vienna, Austria

[2] Faculty of Computer Science, Free University of Bozen-Bolzano, Bolzano, Italy

[3] Vienna University of Technology, Favoritenstraße 9, 1040 Vienna, Austria

# Updating RDFS ABoxes and TBoxes in SPARQL

Albin Ahmeti[1], Diego Calvanese[2], and Axel Polleres[3]

[1] Vienna University of Technology, Favoritenstraße 9, 1040 Vienna, Austria
[2] Faculty of Computer Science, Free University of Bozen-Bolzano, Bolzano, Italy
[3] Vienna University of Economics and Business, Welthandelsplatz 1, 1020 Vienna, Austria

**Abstract.** Updates in RDF stores have recently been standardised in the SPARQL 1.1 Update specification. However, computing answers entailed by ontologies in triple stores is usually treated orthogonal to updates. Even the W3C's recent SPARQL 1.1 Update language and SPARQL 1.1 Entailment Regimes specifications explicitly exclude a standard behaviour how SPARQL endpoints should treat entailment regimes other than simple entailment in the context of updates. In this paper, we take a first step to close this gap. We define a fragment of SPARQL basic graph patterns corresponding to (the RDFS fragment of) *DL-Lite* and the corresponding SPARQL update language, dealing with updates both of ABox and of TBox statements. We discuss possible semantics along with potential strategies for implementing them. We treat both, (i) materialised RDF stores, which store all entailed triples explicitly, and (ii) reduced RDF Stores, that is, redundancy-free RDF stores that do not store any RDF triples (corresponding to *DL-Lite* ABox statements) entailed by others already.

## 1 Introduction

SPARQL provides a standard for accessing structured Data on the Web and the availability of such a standard may well be called one of the key factors to the success and increasing adoption of RDF and the Semantic Web. Still, in its first iteration the SPARQL [24] specification has neither defined any standard way how to treat ontological entailments with respect to RDF Schema (RDFS) and OWL ontologies, nor provided standard means how to update dynamic RDF data. Both these gaps have been addressed within the recent SPARQL 1.1 specification, which both provides means to define query answers under ontological entailments (SPARQL 1.1 Entailment Regimes [9]) as well as an update language to update RDF data stored in a triple store (SPARQL 1.1 Update [8]). Nonetheless, these specifications do not provide a standard behaviour how SPARQL endpoints should treat entailment regimes other than simple entailment in the context of updates. The main issue of updates under entailments is how updates shall deal with implied statements:

- What does it mean if an implied triple is explicitly (re-)inserted (or deleted)?
- Which (if any) additional triples should be inserted, (or, resp., deleted) upon updates?

For the sake of this paper, we address such questions with the focus on a deliberately minimal ontology language, namely the minimal RDFS fragment of [19].[4] As it turns out, even in this confined setting, updates as defined in the SPARQL 1.1 Update specification impose non-trivial challenges; in particular, specific issues arise through the interplay of INSERT, DELETE, and WHERE clauses within a single SPARQL update operation, which —to the best of our knowledge— have not yet been considered in the literature in combination.

---

[4] We ignore issues like axiomatic triples [13], blank nodes [17], or, in the context of OWL, inconsistencies arising through updates [5]. We neither consider named graphs in SPARQL, which is why we talk about "triple stores" as opposed to "graph stores" [8].

**Table 1.** *DL-Lite*$_{\text{RDFS}}$ assertions vs. RDF(S), where $A$, $A'$ denote concept (or, class) names, $P$, $P'$ denote role (or, property) names, $\Gamma$ is a set of constants, and $x, y \in \Gamma$.

| | *DL-Lite*$_{\text{RDFS}}$ TBox | RDFS |
|---|---|---|
| 1 | $A' \sqsubseteq A$ | $A'$ rdfs:subClassOf $A$. |
| 2 | $\exists P \sqsubseteq A$ | $P$ rdfs:domain $A$. |
| 3 | $\exists P^- \sqsubseteq A$ | $P$ rdfs:range $A$. |
| 4 | $P' \sqsubseteq P$ | $P'$ rdfs:subPropertyOf $P$. |

| | *DL-Lite*$_{\text{RDFS}}$ ABox | RDFS |
|---|---|---|
| 5 | $A(x)$ | $x$ rdf:type $A$. |
| 6 | $P(x,y)$ | $x$ P $y$. |

*Example 1.* As a running example, let us assume an RDF triple store $G$, containing RDF data about family relationships (in Turtle [2] syntax)

```
:joe :hasParent :jack. :joe :hasMother :jane.
```

along with an RDFS ontology $O_{fam}$ as follows

```
:hasFather rdfs:subPropertyOf :hasParent.
:hasMother rdfs:subPropertyOf :hasParent.
:Father rdfs:subClassOf :Parent.
:Mother rdfs:subClassOf :Parent.
:hasFather rdfs:range :Father; rdfs:domain :Child.
:hasMother rdfs:range :Mother; rdfs:domain :Child.
:hasParent rdfs:range :Parent; rdfs:domain :Child.
```

The following query returns both :jack and :jane as (RDFS entailed) answers:
**SELECT** ?Y **WHERE {** :joe :hasParent ?Y. **}**
A SPARQL endpoint that only considers simple entailment though, would only return :jack as an answer. ∎

The intended behaviour for the query in Ex. 1 is typically achieved by either *(i)* computing entailed answers at query run-time, which can actually be achieved by query rewriting techniques, or *(ii)* by materialising all implied triples in the store, normally at loading time.

That is, on the one hand, borrowing from query rewriting techniques from *DL-Lite* (such as, e.g., *PerfectRef* [4], several refinements of which have been proposed in the literature[5]) one can reformulate such a query to return also implied answers. While the rewritten query is exponential in the worst case wrt. the length of the original query (and polynomial in the size of the TBox), for moderate size TBoxes and queries, such a rewriting approach is quite feasible.

---
[5] We show in Alg. 1 in the appendix a down-stripped version of *PerfectRef* [4], covering the essentials of RDFS.

*Example 2 (cont'd).* If we interpret the query in Ex. 1 as a conjunctive query, and the RDFS assertions from $O_{fam}$ as a DL TBox, the resulting UCQ as from Alg. 1 can be just translated back into SPARQL as follows.

```
SELECT ?Y WHERE { { :joe :hasParent ?Y. }
            UNION { :joe :hasFather ?Y. }
            UNION { :joe :hasMother ?Y. }}
```

Indeed, this query returns both :jane and :jack. ∎

On the other hand, an alternative[6] is to materialize all inferences in the triple store, such that the original query can be used 'as is', for instance using the minimalistic inference rules for RDFS from [19][7] shown in Fig. 1.

*Example 3 (cont'd).* The materialised version of $G$ would contain the following triples – for conciseness we only show assertional implied triples here, that is triples from the first four rules in Fig. 1.

```
:joe a :Child; :hasParent :jack;
     :hasMother :jane; :hasParent :jane.
:jack a :Parent. :jane a :Mother, :Parent.
```

Obviously, on a materialised triple store, the query from Ex. 1 would already return the expected results. ∎

However, when it comes to *updating* the RDF store in terms of SPARQL 1.1 Update, things become less clear.

*Example 4 (cont'd).* The following operation tries to delete an implied triple and at the same time to (re-)insert another implied triple.

```
DELETE {?X a :Child.}
INSERT {?Y a :Mother.}
WHERE {?X :hasMother ?Y.}
```
∎

Existing triple stores offer different solutions to these problems, ranging from ignoring entailments in updates altogether, to keeping explicit and implicit triples separate and re-materialising upon updates. In the former case (ignoring entailments) updates only refer to explicitly asserted triples, which may result in non-intuitive behaviours, whereas

---

[6] This alternative is viable for RDFS, but not necessarily for more expressive DLs.
[7] These rules correspond to rules 2), 3), 4) from [19]; they suffice since we do not consider blank nodes.

$$\frac{?P \text{ rdfs:subPropertyOf } ?Q. \quad ?S \ ?P \ ?O.}{?S \ ?Q \ ?O.} \qquad \frac{?P \text{ rdfs:range } ?C. \quad ?S \ ?P \ ?O.}{?O \text{ rdf:type } ?C.}$$

$$\frac{?C \text{ rdfs:subClassOf } ?D. \quad ?S \text{ rdf:type } ?C.}{?S \text{ rdf:type } ?D.} \qquad \frac{?P \text{ rdfs:subPropertyOf } ?Q. \quad ?Q \text{ rdfs:subPropertyOf } ?R.}{?P \text{ rdfs:subPropertyOf } ?R.}$$

$$\frac{?P \text{ rdfs:domain } ?C. \quad ?S \ ?P \ ?O.}{?S \text{ rdf:type } ?C.} \qquad \frac{?C \text{ rdfs:subClassOf } ?D. \quad ?D \text{ rdfs:subClassOf } ?E.}{?C \text{ rdfs:subClassOf } ?E.}$$

**Fig. 1.** Minimal RDFS inference rules

the latter case (re-materialisation) may be very costly – while still not eliminating all non-intuitive cases, as we will see. The problem is aggravated by no systematic approach to which implied triples to store explicitly in a triple store and which not. In this paper we try to argue for a more systematic approach for dealing with updates in the context of RDFS entailments. More specifically, we will distinguish between two kinds of triple stores, that is *(i) materialised RDF stores*, which store all entailed ABox triples explicitly, and *(ii) reduced RDF Stores*, that is, redundancy-free RDF stores that do not store any assertional (ABox) triples entailed by others already. We propose alternative update semantics that preserve the respective types *(i)* and *(ii)* of triple stores, and discuss possible implementation strategies, partially inspired by query rewriting techniques from ontology-based data access (OBDA) [15] and *DL-Lite* [4].

As already shown in [11], erasure of ABox statements is deterministic in the context of RDFS, but insertion and particularly the interplay of DELETE/INSERT in SPARQL 1.1 Update has not been considered therein.

The remainder of the paper continues with preliminaries (RDFS, SPARQL, *DL-Lite*, SPARQL update operations) in Sec. 2. We introduce alternative update semantics for materialised and reduced triple stores in Sec. 3, and discuss them in Sec. 4 and Sec. 5, respectively. In Sec. 6 we relax the initial assumption that terminological statements (using the RDFS vocabulary) are static and discuss terminological updates. After a discussion on future and related work in Sec. 7, we conclude in Sec. 8.

## 2 Preliminaries

We introduce some basic notions about RDF graphs, RDFS ontologies, and SPARQL queries. Since we will draw from ideas coming from OBDA and *DL-Lite*, we introduce these notions in a way that is compatible with DLs.

**Definition 1 (RDFS ontology, ABox, TBox, triple store).** *We call a set $\mathcal{T}$ of inclusion assertions of the forms 1–4 in Table 1 an* RDFS ontology, *or* (RDFS) TBox, *a set $\mathcal{A}$ of assertions of the forms 5–6 in Table 1 an* (RDF) ABox, *and the union $G = \mathcal{T} \cup \mathcal{A}$ an* (RDFS) triple store.

In the context of RDF(S), the set $\Gamma$ of constants coincides with the set $I$ of IRIs. We assume the IRIs used for concepts, roles, and constants to be disjoint from IRIs of the RDFS and OWL vocabularies.[8] In the following, we view RDF and DL notation interchangeably, i.e., we treat any RDF graph consisting of triples without non-standard RDFS vocabulary as a set of TBox and ABox assertions. To define the semantics of RDFS, we rely on the standard notions of (first-order logic) *interpretation*, satisfaction of assertions, and *model* (cf. Def. 14 in the appendix).

We now turn to queries, and again treat the cases of SPARQL and DLs interchangeably. Let $\mathcal{V}$ be a countably infinite set of variables (written as alphanumeric strings preceded by '?').

---

[8] That is, we assume no "non-standard use" [23] of these vocabularies. While we could assume concept names, role names and individual constants mutually disjoint, we rather distinguish implicitly between them "per use" (in the sense of "punning" [18]) based on their position in atoms or RDF triples.

**Definition 2 (BGP, CQ, UCQ).** *A* conjunctive query *(CQ) q, or* basic graph pattern *(BGP), is a set of atoms of the forms 5–6 from Table 1, where now $x, y \in \Gamma \cup \mathcal{V}$. A* union of conjunctive queries *(UCQ) Q, or* UNION *pattern, is a set of CQs. We denote with $\mathcal{V}(q)$ (or $\mathcal{V}(Q)$) the set of variables from $\mathcal{V}$ occurring in q (resp., Q).*

Notice that in this definition we are considering only CQs in which all variables are *distinguished* (i.e., are answer variables), and that such queries correspond to SPARQL basic graph patterns (BGPs). From the SPARQL perspective, we allow only for restricted forms of general SPARQL BGPs that correspond to standard CQs as formulated over a DL ontology; that is, we rule out on the one hand more complex patterns in SPARQL 1.1 [12] (such as OPTIONAL, NOT EXISTS, FILTER), and queries with variables in predicate positions, and on the other hand "terminological" queries, e.g., {?x rdfs:subClassOf ?y.}. We will relax this latter restriction later (see Sec. 6). Also, we do not consider here blank nodes separately[9]. By these restrictions, we can treat query answering and BGP matching in SPARQL analogously and define it in terms of interpretations and models (as usual in DLs). Specifically, an *answer* (under RDFS Entailment) to a CQ $q$ over a triple store $G$ is a substitution $\theta$ from the variables in $\mathcal{V}(q)$ to constants in $\Gamma$ such that every model of $G$ satisfies all facts in $q\theta$. We denote the set of all such answers with $ans_{\text{rdfs}}(q, G)$ (or simply $ans(q, G)$). The set of answers to a UCQ $Q$ is $\bigcup_{q \in Q} ans(q, G)$.

From now on, let $rewrite(q, \mathcal{T})$ be the UCQ resulting from applying *PerfectRef* (or, equivalently, the down-stripped version in Alg. 1) to a CQ $q$ and a triple store $G = \mathcal{T} \cup \mathcal{A}$, and let $mat(G)$ be the triple store obtained from exhaustive application on $G$ of the inference rules in Fig. 1.

The next result follow immediately from, e.g., [4, 11, 19] and shows that query answering under RDF can be done by either query rewriting or materialisation.

**Proposition 1.** *Let $G = \mathcal{T} \cup \mathcal{A}$ be a triple store, $q$ a CQ, and $\mathcal{A}'$ the set of ABox assertions in $mat(G)$. Then, $ans(q, G) = ans(rewrite(q, \mathcal{T}), \mathcal{A}) = ans(q, \mathcal{A}')$.*

Various existing triple stores (e.g., BigOWLIM [3]) have the option to perform ABox materialisation directly upon loading data. On the other hand these triple stores do not necessarily materialise the TBox: in order to correctly answer UCQs as defined above, a triple store actually does not need to consider the last two rules in Fig. 1. Accordingly, we will call a triple store or (ABox) *materialised* if in each state it always guarantees $G \setminus \mathcal{T} = mat(G) \setminus mat(\mathcal{T})$. On the other extreme, we find triple stores that do not store *any* redundant ABox triples. By $red(G)$ we denote the hypothetical operator that produces the reduced "core" of $G$, and we call a triple store *(ABox) reduced* if $G = red(G)$. While we do not provide the algorithm to compute $red(G)$, we note that this core is uniquely determined in our setting whenever $\mathcal{T}$ is acyclic (which is usually a safe assumption)[10]; it could be naïvely computed by exhaustively "marking" each triple that can be inferred from applying any of the rules in Fig. 1, and subsequently removing all marked elements of $\mathcal{A}$. We observe that, trivially, a triple store containing no ABox statements is both reduced and materialised.

---

[9] Blank nodes in a triple store may be considered as constants and we do not allow blank nodes in queries, which does not affect the expressivity of SPARQL.

[10] We note that even in the case when the TBox is cyclic we could define a deterministic way to remove redundancies, e.g., by preserving within a cycle only the lexicographically smallest ABox statements. That is, given TBox $A \sqsubseteq B \sqsubseteq C \sqsubseteq A$ and ABox $A(x), C(x)$; we would delete $C(x)$ and retain $A(x)$ only, to preserve reducedness.

Finally, we introduce the notion of a SPARQL update operation.

**Definition 3 (SPARQL update operation).** *Let $P_d$ and $P_i$ be BGPs, and $P_w$ a BGP or UNION pattern. Then an* update operation $u(P_d, P_i, P_w)$ *has the form*

$$\texttt{DELETE} \quad P_d \quad \texttt{INSERT} \quad P_i \quad \texttt{WHERE} \quad P_w$$

Intuitively, the semantics of executing $u(P_d, P_i, P_w)$ on $G$, denoted as $G_{u(P_d, P_i, P_w)}$ is defined by interpreting both $P_d$ and $P_i$ as "templates" to be instantiated with the solutions of $ans(P_w, G)$, resulting in sets of ABox statements $\mathcal{A}_d$ to be deleted from $G$, and $\mathcal{A}_i$ to be inserted into $G$. A naïve update semantics follows straightforwardly.

**Definition 4 (Naïve update semantics).** *Let $G = \mathcal{T} \cup \mathcal{A}$ be a triple store, and $u(P_d, P_i, P_w)$ an update operation. Then,* naive update of $G$ with $u(P_d, P_i, P_w)$*, denoted $G_{u(P_d, P_i, P_w)}$, is defined as $(G \setminus \mathcal{A}_d) \cup \mathcal{A}_i$, where $\mathcal{A}_d = \bigcup_{\theta \in ans(P_w, G)} gr(P_d\theta)$, $\mathcal{A}_i = \bigcup_{\theta \in ans(P_w, G)} gr(P_i\theta)$, and $gr(P)$ denotes the set of ground triples in a pattern $P$.*

As easily seen, this naïve semantics neither preserves reduced nor materialised triple stores. Consider, e.g., the update from Ex. 4, respectively on the reduced triple store from Ex. 1 and on the materialised triple store from Ex. 3.

## 3 Defining Alternative Update Semantics

We investigate now alternative semantics for ABox updates that preserve either materialised or reduced stores, and discuss in how far these semantics can —similar to query answering— be implemented in off-the-shelf SPARQL 1.1 triple stores by simple rewriting techniques. Specifically, we consider different variants of materialised ABox preserving (or simply, *mat-preserving*) semantics and reduced ABox preserving (or simply, *red-preserving*) semantics for $u(P_d, P_i, P_w)$:

**Sem$_0^{mat}$:** As a baseline for a mat-preserving semantics, we apply the naïve semantics, followed by (re-)materialisation of the whole triple store.
**Sem$_1^{mat}$:** An alternative approach for a mat-preserving semantics is to follow the so-called "delete and rederive" algorithm [10] for deletions, that is:
 – delete the instantiations of $P_d$ plus *"dangling" effects*, i.e., effects of deleted triples that after deletion are not implied any longer *by any non-deleted triples*;
 – insert the instantiations of $P_i$ *plus all their effects*.
**Sem$_2^{mat}$:** Another mat-preserving semantics could take a different viewpoint with respect to deletions, following the intention to:
 – delete the instantiations of $P_d$ *plus all their causes*;
 – insert the instantiations of $P_i$ *plus all their effects*.

**Sem**$_3^{mat}$: Finally, a mat-preserving semantics could combine **Sem**$_1^{mat}$ and **Sem**$_2^{mat}$ with the intention to delete both causes of the instantiations of $P_d$ and (recursively) "dangling" effects.[11]

**Sem**$_0^{red}$: Again, the baseline for a red-preserving semantics would be to apply the naïve semantics, followed by (re-)reducing the triple store.

**Sem**$_1^{red}$: This red-preserving semantics extends **Sem**$_0^{red}$ by additionally deleting the causes of instantiations of $P_d$.

The definitions of semantics **Sem**$_0^{mat}$ and **Sem**$_0^{red}$ are straightforward.

**Definition 5 (Baseline mat-preserving and red-preserving update semantics).** *Let $G$ and $u(P_d, P_i, P_w)$ be as in Def. 4. Then, we define* **Sem**$_0^{mat}$ *and* **Sem**$_0^{red}$ *as follows:*

$$G^{\mathbf{Sem}_0^{mat}}_{u(P_d,P_i,P_w)} = mat(G_{u(P_d,P_i,P_w)}) \qquad G^{\mathbf{Sem}_0^{red}}_{u(P_d,P_i,P_w)} = red(G_{u(P_d,P_i,P_w)})$$

Let us proceed with a quick "reality-check" on these two baseline semantics by means of our running example.

*Example 5.* Consider the update from Ex. 4. It is easy to see that neither under **Sem**$_0^{mat}$ executed on the materialised triple store of Ex. 3, nor under **Sem**$_0^{red}$ executed on the reduced triple store of Ex. 1, it would have *any* effect. ∎

This behaviour is quite arguable. Hence, we proceed with discussing the implications of the proposed alternative update semantics, and how they could be implemented.

## 4 Alternative Mat-Preserving Semantics

We consider now in more detail the mat-preserving semantics. As for **Sem**$_1^{mat}$, we rely on a well-known technique in the area of updates for deductive databases called "delete and rederive" (DRed) [6, 10, 16, 26, 27]. Informally translated to our setting, when given a logic program $\Pi$ and its materialisation $T_\Pi^\omega$, plus a set of rules $A_d$ to be deleted and a set of facts $A_i$ to be inserted, DRed *(i)* first deletes $A_d$ and all its effects (computed via semi-naive evaluation [25]) from $T_\Pi^\omega$, resulting in $(T_\Pi^\omega)'$, *(ii)* then, starting from $(T_\Pi^\omega)'$, re-materialises $(\Pi \setminus A_d) \cup A_i$ (again using semi-naive evaluation).

The basic intuition behind DRed of deleting effects of deleted triples and then re-materializing can be expressed in our notation as follows; as we will consider a variant of this semantics later on, we refer to this semantics as **Sem**$_{1a}^{mat}$.

**Definition 6.** *Let $G = \mathcal{T} \cup \mathcal{A}$ be a triple store and $u(P_d, P_i, P_w)$ an update operation. Then*

$$G^{\mathbf{Sem}_{1a}^{mat}}_{u(P_d,P_i,P_w)} = mat(\mathcal{T} \cup (\mathcal{A} \setminus mat(\mathcal{T} \cup \mathcal{A}_d)) \cup \mathcal{A}_i)$$

*where $\mathcal{A}_d$ and $\mathcal{A}_i$ are defined as in Def. 4.*

---

[11] Note the difference to the basic "delete and rederive" approach. **Sem**$_1^{mat}$ in combination with the intention of **Sem**$_2^{mat}$ would also mean to recursively delete effects of causes, and so forth.

As opposed to the classic DRed algorithm, where Datalog distinguishes between view predicates (IDB) and extensional knowledge in the Database (IDB), in our setting we do not make this distinction, i.e., we do not distinguish between implicitly and explicitly inserted triples. This means that $\mathbf{Sem}_{1a}^{mat}$ would delete also those effects that had been inserted explicitly before.

We introduce now a different variant of this semantics, denoted $\mathbf{Sem}_{1b}^{mat}$, that makes a distinction between explicitly and implicitly inserted triples.

**Definition 7.** *Let $u(P_d, P_i, P_w)$ be an update operation, and $G = \mathcal{T} \cup \mathcal{A}_{expl} \cup \mathcal{A}_{impl}$ a triple store, where $\mathcal{A}_{expl}$ and $\mathcal{A}_{impl}$ denote the explicit and implicit ABox triples, respectively. Then*
$$G^{\mathbf{Sem}_{1b}^{mat}}_{u(P_d, P_i, P_w)} = \mathcal{T} \cup \mathcal{A}'_{expl} \cup \mathcal{A}'_{impl}$$
*where $\mathcal{A}_d$ and $\mathcal{A}_i$ are defined as in Def. 4, and*

$$\mathcal{A}'_{expl} = (\mathcal{A}_{expl} \setminus \mathcal{A}_d) \cup \mathcal{A}_i \qquad \mathcal{A}'_{impl} = mat(\mathcal{A}'_{expl} \cup \mathcal{T}) \setminus \mathcal{T}$$

Note that in $\mathbf{Sem}_{1b}^{mat}$, as opposed to $\mathbf{Sem}_{1a}^{mat}$, we do not explicitly delete effects of $\mathcal{A}_d$ from the materialisation, since the definition just relies on re-materialisation from the explicit ABox $\mathcal{A}'_{expl}$ from scratch. Nonetheless, the original DRed algorithm can still be used for computing $\mathbf{Sem}_{1b}^{mat}$ as shown by the following proposition.

**Proposition 2.** *Let us interpret the inference rules in Fig. 1 and triples in G respectively as rules and facts of a logic program $\Pi$; accordingly, we interpret $\mathcal{A}_d$ and $\mathcal{A}_i$ from Def. 7 as facts to be deleted from and inserted into $\Pi$, respectively. Then, the materialisation computed by DRed, as defined in [16], computes exactly $\mathcal{A}'_{impl}$.*

Note that none of $\mathbf{Sem}_0^{mat}$, $\mathbf{Sem}_{1a}^{mat}$, and $\mathbf{Sem}_{1b}^{mat}$ is equivalent, as shown by the following example.

*Example 6.* Given the following triple store $G$

```
{ :C rdfs:subclassOf :D .    :D rdfs:subclassOf :E }
```

on which we perform the operation `INSERT{:x a :C, :D, :E.}`, *explicitly* adding three triples, and subsequently perform `DELETE{:x a :C, :E.}`, we obtain, according to the three semantics discussed so far the following ABoxes:
$\mathbf{Sem}_0^{mat}$: `{:x a :D. :x a :E.}`    $\mathbf{Sem}_{1a}^{mat}$: `{}`
$\mathbf{Sem}_{1b}^{mat}$: `{:x a :D. :x a :E.}`
While after this update $\mathbf{Sem}_0^{mat}$ and $\mathbf{Sem}_{1b}^{mat}$ deliver the same result, the difference between these two is shown by the subsequent update `DELETE{:x a :D.}`
$\mathbf{Sem}_0^{mat}$: `{:x a :E.}`    $\mathbf{Sem}_{1a}^{mat}$: `{}`    $\mathbf{Sem}_{1b}^{mat}$: `{}`    ∎

As for the subtle difference between $\mathbf{Sem}_{1a}^{mat}$ and $\mathbf{Sem}_{1b}^{mat}$, we point out that none of [16, 26], who both refer to using DRed in the course of RDF updates, make it clear whether explicit and implicit ABox triples are to be treated differently.

Further, continuing on Ex. 5 above, the update from Ex. 4 still would not have *any* effect, neither using $\mathbf{Sem}_{1a}^{mat}$, nor $\mathbf{Sem}_{1b}^{mat}$. That is, it is not possible in any of the update semantics so far to remove information that is implicitly given (without explicitly removing all its causes).

$\mathbf{Sem}_2^{mat}$ aims at addressing this problem concerning the deletion of implicit information. As it turns out, while the intention of $\mathbf{Sem}_2^{mat}$ to delete causes of deletions cannot be captured just with the $mat$ operator, it can be achieved fairly straightforwardly, building upon ideas similar to those used in query rewriting.

As we have seen, in the setting of RDFS we can use Alg. 1 *rewrite* to expand a CQ to a UCQ that incorporates all its "causes". A slight variation can of course be used to compute the set of all causes, that is, in the most naïve fashion by just "flattening" the set of sets returned by Alg. 1 to a simple set; we denote this flattening operation on a set of sets $S$ as *flatten(S)*.

Likewise, we can easily define a modified version of $mat(G)$, applied to a BGP $P$ using a TBox $\mathcal{T}$[12]. Let us call the resulting algorithm $mat_{eff}(P, \mathcal{T})$[13]. Using these considerations, we can thus define both rewritings that consider all causes, and rewritings that consider all effects of a given (insert or delete) pattern $P$:

**Definition 8 (Cause/Effect rewriting).** *Given a BGP insert or delete template $P$ for an update operation over the triple store $G = \mathcal{T} \cup \mathcal{A}$, we define the* all-causes-rewriting *of $P$ as $P^{caus} = \mathit{flatten}(\mathit{rewrite}(P, \mathcal{T}))$; likewise, we define the* all-effects-rewriting *of $P$ as $P^{eff} = \mathit{mat}_{\mathit{eff}}(P, \mathcal{T})$.*

This leads (almost) straightforwardly to a rewriting-based definition of $\mathbf{Sem}_2^{mat}$:

**Definition 9.** *Let $u(P_d, P_i, P_w)$ be an update operation. Then*

$$G^{\mathbf{Sem}_2^{mat}}_{u(P_d,P_i,P_w)} = G_{u(P_d^{caus}, P_i^{eff}, \{rewrite(P_w)\}\{P_d^{fvars}\})}$$

*where $P_d^{fvars} = \{?x \text{ a rdfs:Resource.} \mid \text{for each } ?x \in \mathit{Var}(P_d^{caus}) \setminus \mathit{Var}(P_d)\}$.*

The only tricky part in this definition is the rewriting of the WHERE clause, where $rewrite(P_w)$ is joined[14] with a new pattern $P_d^{fvars}$ that binds the "free" variables (i.e., the new variables denoted by '_' in Table 2, which were introduced by Alg. 1) in the rewritten DELETE clause, $P_d^{caus}$. Here, `?x a rdfs:Resource.` is a shortcut for the UCQ `{{?x ?x_p ?x_o} UNION {?x_s ?x ?x_o} UNION {?x_s ?x_p ?x}}`, which binds any term occurring in the triple store.

*Example 7.* Getting back to the materialised version of the triple store $G$ from Ex. 3, the update $u$ from Ex. 4 would, according to $\mathbf{Sem}_2^{mat}$, be rewritten to

```
DELETE { ?X a :Child. ?X :hasFather ?x1.
         ?X :hasMother ?x2. ?X :hasParent ?x3. }
INSERT { ?Y a :Mother. ?Y a :Parent. }
```

---

[12] This could be viewed as simply applying the first four inference rules in Fig. 1 exhaustively to $P \cup \mathcal{T}$, and then removing $\mathcal{T}$.

[13] Note that it is not our intention to provide optimized algorithms here, but just to convey the feasibility of this rewriting.

[14] A sequence of '{}'-delimited patterns in SPARQL corresponds to a join, where such joins can again be nested with UNIONs, with the obvious semantics, for details cf. [12].

```
WHERE   { { ?X :hasMother ?Y. }
          { ?x1 a rdfs:Resource.
            ?x2 a rdfs:Resource.
            ?x3 a rdfs:Resource.} }
```
with $G_u^{\mathbf{Sem}_2^{mat}}$ containing the triples :jane a :Mother, :Parent. :jack a :Parent. ∎

It is easy to argue that $\mathbf{Sem}_2^{mat}$ is mat-preserving. However, this semantics might still result in potentially non-intuitive behaviours. For instance, the following subsequent calls of INSERTs and DELETEs might leave "traces", as shown by the following example.

*Example 8.* Assume $G = O_{fam}$ from Ex. 1 with an empty ABox. Under $\mathbf{Sem}_2^{mat}$, the following sequence of updates would leave as a trace —among others— the resulting triples as in Ex. 7, which would not be the case under the naïve semantics.
```
DELETE{} INSERT {:joe :hasMother :jane; :hasFather :jack} WHERE{};
DELETE {:joe :hasMother :jane; :hasFather :jack} INSERT{} WHERE{}
```
∎

$\mathbf{Sem}_3^{mat}$ tries to address the issue of such "traces", but can no longer be formulated by a relatively straightforward rewriting. For the present, preliminary paper we leave out a detailed definition/implementation capturing the intention of $\mathbf{Sem}_3^{mat}$; there are two possible starting points, namely combining $\mathbf{Sem}_{1a}^{mat}+\mathbf{Sem}_2^{mat}$, or $\mathbf{Sem}_{1b}^{mat}+\mathbf{Sem}_2^{mat}$, resp. We emphasise though, that independent of this choice, a semantics that achieves the intention of $\mathbf{Sem}_3^{mat}$ would still potentially run into arguable cases, since it might run into removing seemingly "disconnected" implicit assertions, whenever removed assertions cause these, as shown by the following example.

*Example 9.* Assume a materialised triple store $G$ consisting only of the TBox triples :Father rdfs:subClassOf :Person, :Male . The behaviour of the following update sequence under a semantcs implementing $\mathbf{Sem}_3^{mat}$ intention is arguable:
```
DELETE {} INSERT {:x a :Father.} WHERE {};
DELETE {:x a :Male.} INSERT {} WHERE {};
```
We leave it open for now whether "recursive deletion of dangling effects" is intuitive: in this case, should upon deletion of $x$ being male, also be deleted that $x$ is a Person? ∎

In a strict reading of $\mathbf{Sem}_3^{mat}$'s intention, :x a :Person. would count as a dangling effect of the cause for :x a :Male., since it is an effect of the triple in the insertion with no other causes in the triple store, and thus should be removed upon the delete operation.

Lastly, we point out that while implementations of (materialized) triple stores may make a distinction between implicit and explicitly inserted triples (e.g., by storing explicit and implicit triples separately, as sketched in $\mathbf{Sem}_{1b}^{mat}$ already), we consider the distinction between implicit triples and explicitly inserted ones non-trivial in the context of SPARQL 1.1 Update: for instance, is a triple inserted based upon implicit bindings in the WHERE clause of an INSERT statement to be considered "explicitly inserted" or not? We tend towards avoiding such distinction, but we have more in-depth discussions of such philosophical aspects of possible SPARQL update semantics on our agenda; For now, we turn our attention to the potential alternatives for red-preserving semantics.

## 5 Alternative Red-Preserving Semantics

Again, similar to $\mathbf{Sem}_3^{mat}$, for both baseline semantics $\mathbf{Sem}_0^{red}$ and $\mathbf{Sem}_1^{red}$ we leave it open whether they can be implemented by rewriting to SPARQL update operations following the naïve semantics, i.e., without the need to apply $red(G)$ over the whole triple store after each update; a strategy to avoid calling $red(G)$ would roughly include the following steps:

– delete the instantiations $P_d$ *plus all the effects of instantiations* of $P_i$, which will be implied anyway upon the new insertion, thus preserving reduced;
– insert instantiations of $P_i$ only if they are *not implied*, that is, they are not already implied by the current state of $G$ or all their causes in $G$ were to be deleted.

We leave further investigation of whether these steps can be cast into update requests directly by rewriting techniques to future work. Rather, we show that we can capture the intention of $\mathbf{Sem}_1^{red}$ by a straightforward extension of the baseline semantics.

**Definition 10 ($\mathbf{Sem}_1^{red}$).** *Let $u(P_d, P_i, P_w)$ be an update operation. Then*

$$G_{u(P_d,P_i,P_w)}^{\mathbf{Sem}_1^{red}} = red(G_{u(P_d^{caus},P_i,\{rewrite(P_w)\}\{P_d^{fvars}\})})$$

*where $P_d^{caus}$ and $P_d^{fvars}$ are as before.*

*Example 10.* Getting back to the reduced version of the triple store $G$ from Ex. 1, the update $u$ from Ex. 4 would, according to $\mathbf{Sem}_1^{red}$, be rewritten to
```
DELETE { ?X a :Child. ?X :hasFather ?x1.
         ?X :hasMother ?x2. ?X :hasParent ?x3. }
INSERT { ?Y a :Mother. }
WHERE  { { ?X :hasMother ?Y. }
         { ?x1 a rdfs:Resource.
           ?x2 a rdfs:Resource.
           ?x3 a rdfs:Resource.} }
```
with $G_u^{\mathbf{Sem}_1^{red}}$ containing the triple :jane a :Mother.. Observe here the deletion of the triple :joe :hasParent :jack., which some might view as non-intuitive. ∎

In a reduced store effects of $P_d$ need not be deleted, which makes the considerations that lead us to $\mathbf{Sem}_3^{mat}$ irrelevant for a red-preserving semantics, as shown next.

*Example 11.* Under $\mathbf{Sem}_1^{red}$, as opposed to $\mathbf{Sem}_2^{mat}$, the update sequence of Ex. 8 would leave no traces. However, the update sequence of Ex. 9 would likewise result in an empty ABox, again losing idempotence of single triple insertion followed by deletion. ∎

Note that whereas the rewriting for $\mathbf{Sem}_1^{red}$ is similar to that for $\mathbf{Sem}_2^{mat}$, post-processing for preserving reducedness is not available in off-the-shelf triple stores. Instead, $\mathbf{Sem}_2^{mat}$ could be readily executed by rewriting on existing triple stores, preserving materialization.

## 6 TBox Updates

So far, we have considered the TBox as static. As already noted in [11], additionally allowing TBox updates considerably complicates issues and opens various more degrees of freedom for possible semantics. While it is out of the scope of this paper to explore all of these, we limit ourselves to sketch these different degrees of freedom and suggest one pragmatic approach to extend updates expressed in SPARQL to RDFS TBoxes.

In order to allow for TBox updates, we have to extend the update language: in the following, we will assume *general BGPs*, extending Def. 2.

**Definition 11 (general BGP).** *A* general BGP *is a set of triples of any of the forms from Table 1, where $x, y, A, A', P, P' \in \Gamma \cup \mathcal{V}$.*

We observe that with this relaxation for BGPs, updates as per Def. 3 can query TBox data, since they admit TBox triples in $P_w$. In order to address this issue we need to also generalize the definition of query answers.[15]

**Definition 12.** *Let $Q$ be the union of general BGPs and $[[Q]]_G$ the simple SPARQL semantics as per [21], i.e., essentially the set of query answers obtained as the union of answers from simple pattern matching of the general BGPs in $Q$ over the graph $G$. Then we define $ans_{RDFS}(Q, G) = [[Q]]_{mat(G)}$*

In fact, Def. 12 does not affect ABox inferences, that is, the following corollary follows immediately from Prop. 1.

**Corollary 1.** *Let $Q$ be a (non-general) UCQ according to Def. 2. Then $ans_{RDFS}(Q, G) = ans_{rdfs}(Q, G)$*

However, as opposed to the setting discussed so far, where the last two rules in Fig. 1 used for TBox materialisation did not play a significant role, these rules are essential for completeness of terminological queries. In this context, we also note that upon such terminological queries, the RDFS semantics and DL semantics differ, since the standard "intensional" semantics of RDFS, which is essentially defined by the inference rules in Fig. 1, does not cover all terminological inferences that would be derivable in DL. A recent paper [7] provides an extended set of "extensional" inference rules to close this gap, but we leave a more detailed discussion to future work. Instead, we focus on the discussion of updates under the standard "intensional" semantics for now and attempt to define a reasonable (that means computable) semantics under this setting.

*Observation 1.* TBox updates potentially affect both materialization and reducedness of the ABox, that is, *(i)* upon TBox *insertions* a materialized ABox might need to be re-materialized in order to preserve materialization, and, respectively, a reduced ABox might no longer be reduced; *(ii)* upon TBox *deletions* in a materialized setting, we have a similar issue to what we called "dangling" effects, whereas in a reduced setting indirect deletions of implied triples could cause unintuitive behaviour.

*Observation 2.* Whereas deletions of implicit ABox triples can be achieved deterministically by deleting all single causes, TBox deletions involving `rdfs:subClassOf` `rdfs:subPropertyOf` chains can be achieved in several distinct ways, as already observed by [11].

---

[15] As mentioned in Fn. 8, elements of $\Gamma$ may act as individuals, concept, or roles names in parallel.

*Example 12.* Let us assume the following graph $G$

```
:A rdfs:subClassOf :B . :B rdfs:subClassOf :C .
:B rdfs:subClassOf :D . :C rdfs:subClassOf :E .
:D rdfs:subClassOf :E . :E rfds:subClassOf :F .
```

with the update   `DELETE :A rdfs:subclassOf :F.`

Independent of whether we assume a materialised TBox or not, we would have various choices here on triples to remove to delete all the causes for the TBox statement to be deleted.  ∎

In order to define a deterministic semantics for TBox updates, we need a canonical way to delete implicit and explicit TBox triples. [11] suggest minimal cuts, in the `rdfs:subClassOf` (or `rdfs:subPropertyOf`, resp.) graphs as candidates for deletions of `rdfs:subClassOf` (or `rdfs:subPropertyOf`, resp.) triples. However, as easily verified by Ex. 12, minimal multicuts are still ambiguous.

Here, we suggest two update semantics using rewritings to SPARQL1.1 property path patterns [12] that yield canonical cuts.

**Definition 13.** *Let $u(P_d, P_i, P_w)$ be an update operation where $P_d$, $P_i$, $P_w$ are general BGPs. Then*
$$G^{\mathbf{Sem}^{mat}_{outcut}}_{u(P_d, P_i, P_w)} = mat(G_{u(P'_d, P_i, P'_w)})$$
*where each triple $\{A_1\ sc\ A_2\} \in P_d$ such that $sc \in \{$rdfs:subClassOf, rdfs:subPropertyOf$\}$ is replaced within $P'_d$ by $\{A_1\ sc\ ?x.\}$, and we add to $P'_w$ the property path pattern $\{A_1\ sc?x.\ ?x\ sc^*\ A_2\}$. Analogously, $\mathbf{Sem}^{mat}_{incut}$ is defined by replacing $\{?x\ sc\ A_2\}$ within $P'_d$, and adding $\{A_1\ sc^*?x.\ ?x\ sc\ A_2\}$ within $P'_w$ instead.*

Both $\mathbf{Sem}^{mat}_{outcut}$ and $\mathbf{Sem}^{mat}_{incut}$ may be viewed as straightforward extensions of $\mathbf{Sem}^{mat}_0$, i.e., both are mat-preserving and equivalent to the baseline semantics for non-general BGPs (i.e., on ABox updates):

**Proposition 3.** *Let $u(P_d, P_i, P_w)$ be an update operation, where $P_d, P_i, P_w$ are (non-general) BGPs, then*

$$G^{\mathbf{Sem}^{mat}_{outcut}}_{u(P_d, P_i, P_w)} = G^{\mathbf{Sem}^{mat}_{incut}}_{u(P_d, P_i, P_w)} = G^{\mathbf{Sem}^{mat}_0}_{u(P_d, P_i, P_w)}$$

The intuition behind the rewriting in $\mathbf{Sem}^{mat}_{outcut}$ is to delete for every deleted $A\ sc\ B$. triple, all directly outgoing $sc$ edges from $A$ that lead into paths to $B$, or, resp., in $\mathbf{Sem}^{mat}_{incut}$ all directly incoming edges to $B$.

We note that the intuition to choose these canonical cuts is motivated by the following proposition.

**Proposition 4.** *Let $u = $ `DELETE` $\{A\ sc\ B\}$, and $G$ a triple store with materialised TBox $\mathcal{T}$. Then, the TBox statement deleted by $G^{\mathbf{Sem}^{mat}_{outcut}}_{u(P_d, P_i, P_w)}$ (or, $G^{\mathbf{Sem}^{mat}_{incut}}_{u(P_d, P_i, P_w)}$, resp.) forms a minimal cut [11] of $\mathcal{T}$ disconnecting $A$ and $B$.*

*Proof (Sketch).* In a materialised TBox, one can reach $B$ from $A$ either directly or via $n$ direct neighbours $C_i \neq B$, which (in)directly connect to $B$. So, a minimal cut contains either the multicut between $A$ and the $C_i$s, or between the $C_i$s and $B$; the latter multicut requires at least the same amount of edges to be deleted as the former — which in turn corresponds to the outbound cut. This proofs the claim for $\mathbf{Sem}_{outcut}^{mat}$. We can proceed analogously for $\mathbf{Sem}_{incut}^{mat}$. □

The following example illustrates that the generalisation of Prop. 4 to updates involving the deletion of several TBox statements at once does not hold.

*Example 13.* Assume the following materialised triple store $G = \{$`:A` $sc$ `:B,:C,:D.` `:B` $sc$ `:C, :D.`$\}$ and $u =$ `DELETE{:A` $sc$ `:C. :A` $sc$ `:D.}`. Here, $\mathbf{Sem}_{incut}^{mat}$ would not provide a minimal *multicut* in $G$ wrt disconnecting (`:A, :C`) and (`:A, :D`).[16] ∎

As the example shows, the extension of the baseline ABox update semantics to TBox updates already yields new degrees of freedom. We leave a more in-depth discussion of TBox updates also extending the other semantics from Sec. 3 for future work.

## 7 Further Related Work and Possible Future Directions

Previous work on updates in the context of tractable ontologies such as RDFS [11] and *DL-Lite* [5] typically has treated DELETEs and INSERTs in isolation, but not both at the same time nor in combination with templates filled by WHERE clauses, as in SPARQL 1.1; that is, these approaches are not based on BGP matching but rather on a set of ABox assertions to be updated known a priori. Pairing both DELETE and INSERT, as in our case, poses new challenges, which we tried to address by introducing different semantics and taking into account both materialised and reduced stores. In the future, we plan to extend our work in the context of *DL-Lite* where we could build upon, thoroughly studied query rewriting techniques, and at the same time benefiting from a more expressive ontology language. Expanding beyond our simple minimal RDFS language towards more features of *DL-Lite* or coverage of unrestricted RDF graphs would impose new challenges: for instance, consistency checking and consistency-preserving updates (as those treated in [5]) do not yet play a role in the setting of RDFS; extensions in these directions as well as practically evaluating the proposed semantics on existing triple stores is on our agenda.

As for further related works, in the context of reduced stores, we refer to [22], where the cost of redundancy elimination under various (rule-based) entailment regimes —including RDFS— is discussed in detail. In the area of database theory, there has been a lot of work on updating logical databases: Winslett [28] distinguishes between model-based and formula-based updates; our approach clearly falls in the latter category, more concretely, ABox updates could be viewed as sets of propositional knowledge base updates [14] generated by SPARQL instantiating DELETE/INSERT templates. Let us further note that in the more applied area of databases, there are obvious parallels between some of our considerations and CASCADE DELETEs in SQL (that is, deletions under foreign key constraints), in the sense that we trigger additional deletions of causes/effects in some of the proposed updated semantics discussed herein.

---

[16] An orthogonal example, where $\mathbf{Sem}_{outcut}^{mat}$ would not yield a minimal multicut can be constructed symmetrically.

## 8 Conclusions

We have discussed the semantics of SPARQL 1.1 Update in the context of RDFS. To the best of our knowledge, this is the first work to discuss how to combine RDFS with the new SPARQL 1.1 Update language. While we have been operating on a very restricted setting in this paper, we could demonstrate that even in this setting (only capturing minimal RDFS entailments, restricting BGPs to disallow non-standard-use of the RDFS vocabulary) the definition of a SPARQL 1.1 Update semantics under entailments is a non-trivial task. We proposed several possible semantics, neither of which might seem intuitive for all possible use cases; this might well suggest that there is no "one-size-fits-all" update semantics. Further, while ontologies should be "ready for evolution" [20], we believe that more work into semantics for updates of ontologies alongside with data (TBox & ABox) is still needed to ground research in *Ontology Evolution* into standards (SPARQL, RDF, RDFS,OWL), particularly in the light of the increased importance that RDF and SPARQL are experiencing in dynamic domains where also data is continuously updated (dealing with dynamics in Linked Data, querying sensor data or stream reasoning). We hope that to have taken a first step in the present paper.

## A  Appendix

**Definition 14 (Interpretation, satisfaction, model).** *An* interpretation $\langle \Delta^{\mathcal{I}}, \cdot^{\mathcal{I}} \rangle$ *consists of a non-empty set $\Delta^{\mathcal{I}}$ called the object domain, and an interpretation function $\cdot^{\mathcal{I}}$, which maps*
  - *each atomic concept A to a subset $A^{\mathcal{I}}$ of $\Delta^{\mathcal{I}}$,*
  - *each atomic role P to a binary relation $P^{\mathcal{I}}$ over $\Delta^{\mathcal{I}}$, and*
  - *each element of $\Gamma$ to an element of $\Delta^{\mathcal{I}}$.*

*For concept expressions, the interpretation function is defined as follows:*
  - $\exists P^{\mathcal{I}} = \{x \in \Delta^{\mathcal{I}} \mid \exists y.(x,y) \in P^{\mathcal{I}}\}$
  - $(\exists P^-)^{\mathcal{I}} = \{y \in \Delta^{\mathcal{I}} \mid \exists x.(x,y) \in P^{\mathcal{I}}\}$

*An interpretation $\mathcal{I}$* satisfies *an inclusion assertion $E_1 \sqsubseteq E_2$ (of one of the forms 1–4 in Table 1), if $E_1^{\mathcal{I}} \subseteq E_2^{\mathcal{I}}$. Analogously, $\mathcal{I}$* satisfies *an ABox assertion of the form*
  - $A(x)$, *if $x^{\mathcal{I}} \in A^{\mathcal{I}}$, and*
  - $P(x,y)$, *if $(x^{\mathcal{I}}, y^{\mathcal{I}}) \in P^{\mathcal{I}}$.*

*An interpretation $\mathcal{I}$ is called a* model *of a triple store G (resp., a TBox $\mathcal{T}$, an ABox $\mathcal{A}$), denoted $\mathcal{I} \models G$ (resp., $\mathcal{I} \models \mathcal{T}$, $\mathcal{I} \models \mathcal{A}$), if $\mathcal{I}$ satisfies all assertions in G (resp., $\mathcal{T}$, $\mathcal{A}$).*

---

**Algorithm 1:** $rewrite(q, \mathcal{T})$

**Input**: Conjunctive query $q$, TBox $\mathcal{T}$
**Output**: Union (set) of conjunctive queries

1  $P := \{q\}$
2  **repeat**
3  $\quad P' := P$
4  $\quad$ **foreach** $q \in P'$ **do**
5  $\quad\quad$ **foreach** $g$ in $q$ **do**    // expansion
6  $\quad\quad\quad$ **foreach** *inclusion assertion I in $\mathcal{T}$* **do**
7  $\quad\quad\quad\quad$ **if** *I is applicable to g* **then**
8  $\quad\quad\quad\quad\quad$ $P := P \cup \{q[g/\operatorname{gr}(g, I)]\}$
9  **until** $P' = P$
10 **return** $P$

---

**Table 2.** Semantics of $\operatorname{gr}(g, I)$ in Alg. 1. In an atom, '_' stands for a "fresh" variable

| $g$ | $I$ | $\operatorname{gr}(g/I)$ |
|---|---|---|
| $A(x)$ | $A' \sqsubseteq A$ | $A'(x)$ |
| $A(x)$ | $\exists P \sqsubseteq A$ | $P(x, \_)$ |
| $A(x)$ | $\exists P^- \sqsubseteq A$ | $P(\_, x)$ |
| $P(x,y)$ | $P' \sqsubseteq P$ | $P'(x,y)$ |